# How the material world of the brain, as a complex system governed by intermittent dynamics, gives rise to the supposedly "immaterial" world of mental processes such as thought

## Yiannis F. Contoyiannis[1] and Alexandros K. Karlis[2]

1. Department of Electrical and Electronics Engineering, University of West Attica, Ancient Olive Grove Campus, 12241 Egaleo, Greece.

   Visiting Researcher

e-mail: yiaconto@uniwa.gr

2. Department of Physics, National and Kapodistrian University of Athens, GR-15771 Athens, Greece

   Research Collaborator

e-mail: akkalirs@phys.uoa.gr

## Abstract:

In this paper, we argue that the brain is a complex dynamical system in a critical state similar to that of the 3D Ising model. We highlight several properties of the brain, including resonance phenomena within this critical state, long-range autocorrelations consistent with long memory, a hysteresis zone suggesting the presence of an as yet undetected "exotic" type of matter within the brain, and possible quantum phenomena within the microtubules of neural networks. The latter suggests that these structures exhibit a combined form of "logical superposition". In this model, an "AND" rule would require the simultaneous occupation of two distinct quantum states. The proposed "OR" rule allows any one of several quantum states to be occupied at a time. Moreover, we propose that kink and antikink solitons could mediate coupling between occupied quantum states in neuronal microtubules and thereby contribute to pattern formation in neural networks. This mechanism may well be what gives the **brain the flexibility to support complex thought and other cognitive functions.**

# Section 1: Introduction

The brain is the most prominent organ in living organisms. It coordinates the functioning of the entire body and possesses a unique property: the generation of mental functions, namely thought, ideas, imagination, emotions and, ultimately, consciousness. These higher cognitive functions grant the brain self-awareness, leading it to pose the question: "How do I operate?" Since antiquity, we have termed this "immaterial realm" the "Psyche", and all our attempts to interpret it have been metaphysical, entirely removed from science. The present work aims to demonstrate that nothing within our brain is "immaterial". Furthermore, it explores mechanisms that our brain struggles to comprehend because they appear to defy our everyday experience, such as quantum behaviour. Our tools include methodologies that we have developed in recent years, such as the Prime Number Algorithm (PNA) [1,2].

In a previous study [3], we demonstrated that the brain of a healthy individual is in a critical state, in physical terms, whereas the brain of an individual with epilepsy is far from criticality. Furthermore, in a recent series of studies [4,5,6] on finite systems, we identified a hysteresis zone within the critical state that possesses a very rich structure. In the present work, we examine the brain in the light of these findings. Ultimately, we provide comprehensive proof that the healthy brain is indeed in a critical state, in physical terms, and is similar to the 3D Ising spin model in its critical state. We propose the Prime Number Algorithm (PNA) as a mechanism that transforms stimuli and information into the "immaterial" world of the brain's mental processes, such as thought. The main focus of our work is the application of PNA to microscopic biological structures within neurons, known as microtubules. As we demonstrate using PNA, microtubules exhibit quantum behaviour; however, they do not obey fundamental quantum principles such as the superposition of quantum states or quantum entanglement. Doing so would contradict the findings of the present study, which consistently point to the brain being a system in a state of perpetual criticality. The idea of focusing on microtubules was first introduced by Roger Penrose and Stuart Hameroff. Their works articulate the Orch OR (Orchestrated Objective Reduction) theory [7,8,9]. These researchers propose that our brain functions as a quantum computer. In their work [10], N. E. Mavromatos and D. V. Nanopoulos also suggest that microtubules are the microsites within the brain where stable, macroscopic quantum-coherent states emerge. These authors therefore suggest that such features might be important for a model of the brain as a quantum computer.

Our present work takes an entirely different direction, although it also focuses on microtubules. Our approach is that the brain is a system in a critical state, in the physical sense, and that it generates our "mental" world through a unique, semi-quantum process, which we explain in this paper. The ultimate aim of the present work is therefore to investigate how the brain's mental world is material. In other words, we propose a purely physical mechanism by which the brain generates mental processes such as thought, thereby showing that these processes are physical entities rather than an immaterial "Psyche". In neuroscientific terms, thought is a complex, organised chain of electrical and chemical signals across multiple neural networks interconnected through synapses. In this work, we investigate how the network underlying thought is organised in terms of the physics of critical phenomena [11,12].

# Section 2: A brief introduction to second-order phase transitions, the hysteresis zone and resonances in finite-size systems

We have addressed all the topics listed in the title of this section in detail in a series of papers. Because the term "criticality" is used in various ways across mathematics, the social sciences, physics, biology, medicine and economics, we must clarify that we refer to the theory and applications of critical phenomena in physics. A basic property of all systems in nature is that, in a critical state in the physical sense, they obey intermittent dynamics, as described by the theory of deterministic chaos for dynamical systems [13].

$$\phi_{n+1} = \phi_n + u \bullet \phi_n^z + \varepsilon_n, \quad (1)$$

where $\phi_n$ is the $n$th sample of the scaled order parameter, $u > 0$ is a coupling parameter, $z$ denotes a characteristic exponent related to the critical isothermal exponent $\delta$ by

$$z = \delta + 1 \quad (2)$$

and $\varepsilon_n$ denotes the non-universal stochastic noise necessary for the creation of ergodicity [14].

Intermittency consists of successive intervals of low fluctuations, called laminar lengths, interrupted chaotically by intervals of high fluctuations, called bursts. For an order-parameter time series $\phi$ , the laminar lengths correspond to waiting times, i.e., the number of consecutive time-series values that lie within the interval $\left[\phi_0, \phi_L\right]$ (the "laminar region"), bounded by the repelling fixed point $\phi_0$ and by the endpoint of the laminar region, denoted by $\phi_L$. Thus, laminar lengths are estimated within the laminar regions $[\phi_0, \phi_L]$, i.e.,

$$\phi_0 < \phi < \phi_L \quad (3)$$

The dynamics of the fluctuations of the order parameter $\phi$ at the critical state can be determined analytically for a large class of complex systems by introducing the so-called critical intermittent map. This can be approximated by the type-I intermittency map [15,16] given in Eq. (1).

The distribution of appropriately defined waiting times (laminar lengths, as described above) follows a power law of the form [13]:

$$P(L) \sim L^{-p}, \ L>>1. \quad (4)$$

where

$$p = \frac{z}{z-1} \quad (5)$$

As we have shown in our work [16], the exponent p is related to the isothermal critical exponent by:

$$p = 1 + 1/\delta. \quad (6)$$

As is known [11], $\infty > \delta > 1$. Thus, in the critical state, p $\in [1, 2)$.

The exact position of the endpoint $\phi_L$ is unclear [16]; it is therefore treated as a variable parameter. We determine the endpoint by finding the position at which the laminar-length distribution most closely follows a power law (Eq. (4)). Moreover, if the exponent p lies in [1,2), we say that the system producing this time series is in a critical state. The procedure described above constitutes the Method of Critical Fluctuations (MCF) [17,18]. In our work [3], we applied this methodology, combined with the appropriate wavelet analysis, to EEG recordings from a healthy individual and an individual with epilepsy. Wavelet analysis is necessary because it ensures the condition L>>1. Furthermore, in a series of studies in recent years [4,5,6], we found that the critical state exhibits a very rich structure in finite systems.

For thermal systems of infinite size, the established theory of spontaneous symmetry breaking (SSB) [19,20,11] in second-order phase transitions can be expressed in terms of the Ginzburg–Landau (G-L) free energy. As explained in Appendix A, for a second-order phase transition we retain the first two terms of Eq. (1A). Thus, the G-L free energy is:

$$U(\phi) = \frac{1}{2} r_0 \phi^2 + \frac{1}{4} u_0 \phi^4, \qquad (7)$$

The coefficients of Eq. (7) satisfy the following conditions: $r_0, u_0 > 0$ for the symmetric phase, i.e., when the temperature (the control parameter) is equal to or greater than the critical temperature ($T=T_c$) (green curve in Fig. 2(a)), whereas $r_0 < 0$, $u_0 > 0$ for the broken-symmetry phase, i.e., for $T<T_c$ (blue and red curves in Fig. 1(b) and Fig. 1(c), respectively).

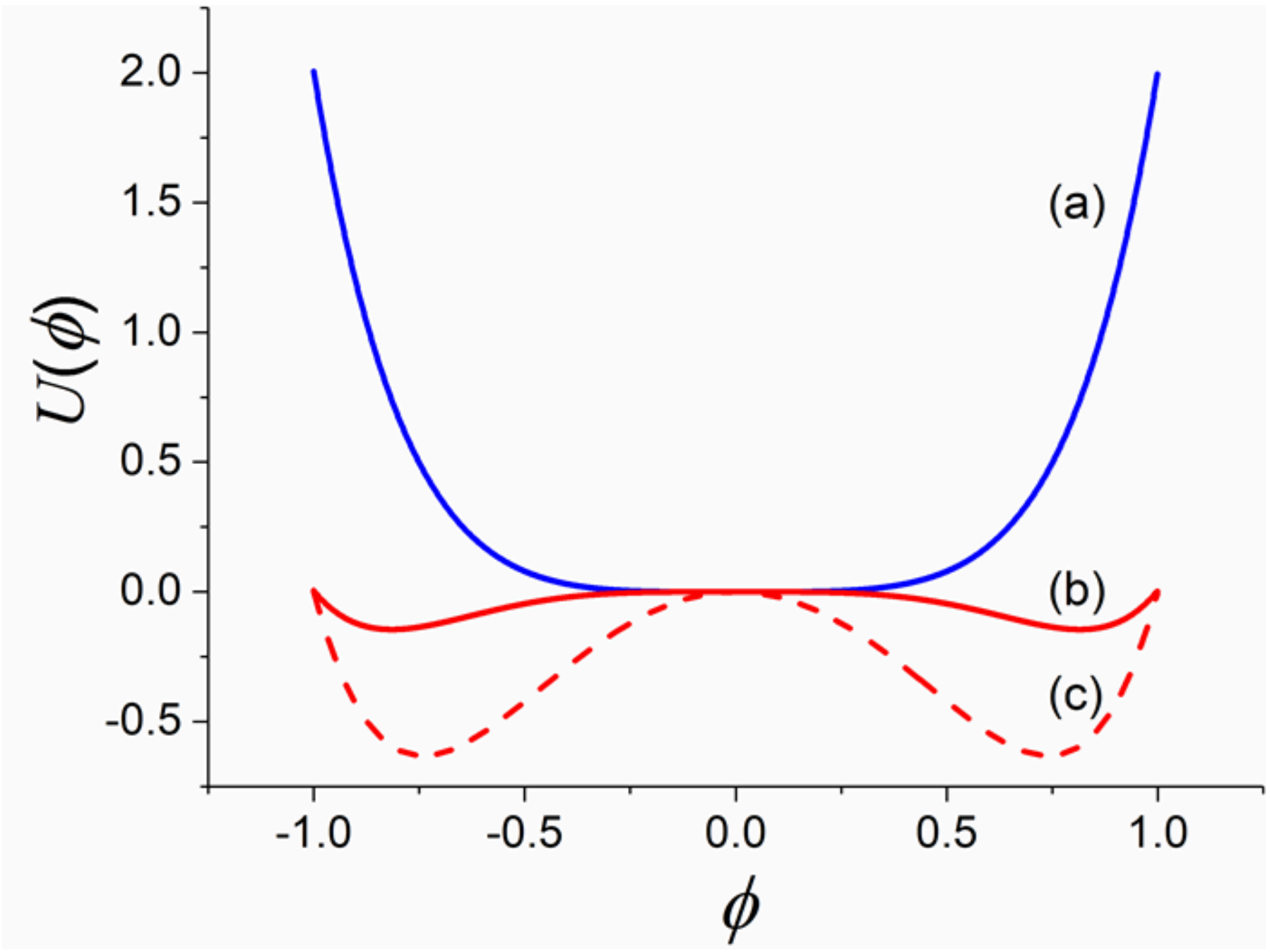

***Fig. 1. The G-L free energy*** $U(\phi)$ ***versus the order parameter*** $\phi$ ***for the second-order transition of a thermal system of infinite size: (a) the symmetric phase (achieved for T≥Tc). (b) SSB (for an arbitrarily selected T<Tc). (c) SSB (for another arbitrarily selected T<Tc).***

As the control parameter (for example, temperature in a thermal system) falls below its critical value, the initial unstable critical point (0,0) ceases to exist and stable vacua appear (red lines in Fig. 1). According to the established theory, the system enters one of the two vacua. This is the process of spontaneous symmetry breaking (SSB). After SSB, a classical system never occupies both vacua simultaneously, because this would involve the quantum principle of superposition of states, which does not apply to classical phase transitions. The above applies to second-order phase transitions in the limit of infinite system size. However, all systems in nature have finite sizes.

The temperature range between the critical (more precisely, pseudocritical) temperature $T_c$ and the SSB temperature $T_{SSB}$ ($T_{SSB}<T_c$), where the critical point appears for the last time, is called the hysteresis zone. Its properties have been studied in detail in our works [2,4]. Fig. A1 in Appendix A presents the distributions of magnetisation values for the 3D Ising model on a $20^3$ lattice. The distributions are shown as the temperature within the zone decreases. Their characteristic feature is the appearance of two lobes that communicate with one another. We begin immediately below the critical temperature and reach the temperature at which the two lobes have almost separated but still coexist for the last time. This is the temperature at which SSB occurs. As shown in [4,5], the width $\Delta\mathrm{T} = T_{pc} - T_{SSB}$ depends on the lattice length through a power law of the form $\Delta T \sim d^{-m}$ (m>1), where d is the lattice length. Thus, the smaller the system, the wider this zone. Consequently, for an infinite system (d→∞), the zone vanishes: ΔT=0. The finite-size approach presented above is therefore in complete agreement with the established theory, i.e., $d \to \infty$ for SSB in G-L $\phi^4$ theory for infinite systems. Within the zone ΔT, three fixed points appear: the initial unstable point (0,0), which acts as a repellor, and two symmetric attractors, which evolve into the stable vacua in Fig. 1 when SSB is complete. Because the initial unstable point (the repellor) persists during this transition, the region is characterised as a hysteresis zone.

We find [6] that the system is in a critical state at all temperatures within the zone. In the interval from T=4.525 to T=4.50, the exponent values show that the criticality belongs to the universality class of the 3D Ising model, for which p=1.20 [16]. From T=4.48 to the SSB temperature T=4.45, the exponent values differ, but all lie within the critical interval [1, 2). At the central value of the hysteresis zone, T=4.49, we find p=1.03, a value very close to the critical limit. The figure below shows the mean laminar length at various temperatures within the zone [6].

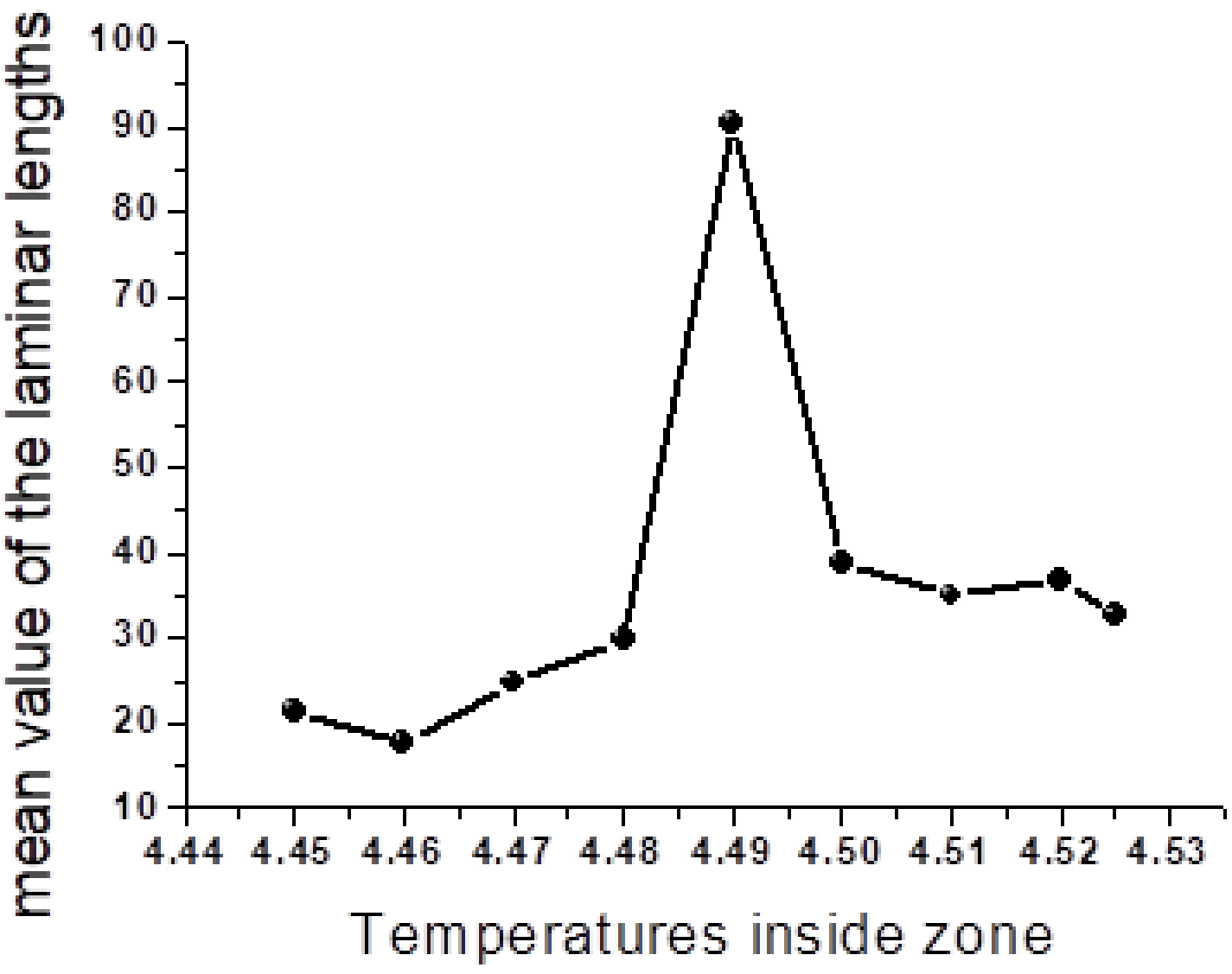

***Fig. 2. The resonance phenomenon within the hysteresis zone. Resonance occurs at the midpoint of the temperature range ΔT, at T=4.49.***

This finding confirms our work [21], in which we proposed, using other examples as well, that the critical state is essentially a state of resonance between all the structures of a complex system. The form of this resonance resembles the plot of the λ-transition phenomenon [12], a second-order phase transition associated particularly with superfluids and superconductors.

# Section 3: A brief presentation of the Haar wavelet transform, which completely ignores noise

An electroencephalogram (EEG) is a medical test that measures electrical activity in the brain and records brainwave patterns. The recording itself is a time series. Although EEG recordings provide information, they also contain substantial noise [22,23,24]. Techniques have been developed to remove this noise. In this section, among other topics, we present an alternative methodology that does not remove the noise but instead ignores its presence entirely.

We have applied this method in numerous previous studies. For completeness, we describe the improved version [25] in detail here. The method is applied to long time series, whether obtained from real-world systems or generated algorithmically. Through the procedure described below, we identify any scaling laws present and calculate the corresponding scaling exponent. If this exponent lies within the critical range [1, 2), it indicates that the system generating the time series is in a critical state. The fundamental property of criticality, as with all complex systems, is self-similarity: the system is "scale-free". This implies that its dynamics encompass all spatio-temporal scales, from the smallest to the largest, and remain invariant under changes of scale. Self-similarity is confirmed if the waiting-time distribution satisfies Eq. (4).

For almost all waiting-time distributions, and particularly for the intermittency considered here, power laws are valid only for L>>1 [26,27,28]. To obtain exact values of the exponents, we must therefore use the new Haar-wavelet-based method, which performs the calculations for L>>1. Its major advantage is that it ignores noise, especially at large scales in noisy experimental data [25]. Thus, large scales carry important information and cannot be excluded from the analysis. The wavelet basis used to develop this algorithm is the Haar wavelet. The wavelet basis elements are $\psi_{j,k}(x) = 2^{\frac{j}{2}}\psi\left(2^j x - k\right)$, with $j = ..., -1, 0, 1, ...$ and $k = - j, ..., -1, 0, 1, ..., j$, whereas the wavelet analysis coefficients are denoted $d_{j,k}$. For $j = k = 0$ one obtains the coarse-grained description of the analysis. Within this description, the analysis coefficients can and do ignore the noise in the signal being analysed [25]. This behaviour has been used to develop an algorithm applicable to any discrete distribution, $P(i)$, $i = 1, 2, ..., N_P$, associated with real or numerical time series. The algorithm can determine whether the distribution being analysed contains power-law behaviour, how closely it follows a power law, and the corresponding exponent $p$. In other words, it provides a form of "fitting" applicable to laminar-length distributions $P(L)$, without the difficulties associated with standard curve-fitting methods due to noise in experimental data, especially at large laminar lengths, where poor tail statistics lead to results of questionable reliability. The mother Haar wavelet is defined using the Heaviside step function, $\Theta(z)$, as:

$$\psi_H(x) = \Theta\left(\frac{\Delta}{2} - x\right)\Theta(x) - \Theta\left(x - \frac{\Delta}{2}\right)\Theta(\Delta - x) \quad (8)$$

for $x \in (0, \Delta]$, while the wavelet analysis coefficients for the expansion of a power-law function (appropriate for a finite system) $F(x) = \{0,\ x \notin [\Delta_{min}, \Delta_{max})\ cx^{-p},\ \Delta_{min} \leq x < \Delta_{max}$ can be written as [25]:

$$d_{j,k} = c\sqrt{\frac{2^j}{\Delta}}\left(\sum_{i=max\left(\left[k\frac{\Delta}{2^j}\right],1\right)}^{\left[k\frac{\Delta}{2^j}+\frac{\Delta}{2^{j+1}}\right]} i^{-p} - \sum_{i=\left[k\frac{\Delta}{2^j}+\frac{\Delta}{2^{j+1}}\right]+1}^{\left[(k+1)\frac{\Delta}{2^j}\right]} i^{-p}\right), \quad (9)$$

where $c$ is a normalisation constant and $[z]$ denotes the integer part of a variable $z$.

One can then define the following quantities [25] for the discrete case:

$$\lambda = \frac{\frac{d_{00}}{d_{10}}}{\frac{d_{10}}{d_{20}}} = \frac{d_{00}d_{20}}{d_{10}^2} = \frac{\left(\sum_{i=1}^{\left[\frac{\Delta}{2}\right]} P(i) - \sum_{\frac{\Delta}{2}}^{\Delta} P(i)\right)\left(\sum_{i=1}^{\left[\frac{\Delta}{8}\right]} P(i) - \sum_{\left[\frac{\Delta}{8}\right]}^{\left[\frac{\Delta}{4}\right]} P(i)\right)}{\left(\sum_{i=1}^{\left[\frac{\Delta}{4}\right]} P(i) - \sum_{\left[\frac{\Delta}{4}\right]}^{\left[\frac{\Delta}{2}\right]} P(i)\right)^2} \quad \textbf{(10)}$$

**and**

$$R = \frac{d_{00}}{d_{10}} = \frac{1}{\sqrt{2}}\frac{\left(\sum_{i=1}^{\left[\frac{\Delta}{2}\right]} P(i) - \sum_{\frac{\Delta}{2}}^{\Delta} P(i)\right)}{\left(\sum_{i=1}^{\left[\frac{\Delta}{4}\right]} P(i) - \sum_{\left[\frac{\Delta}{4}\right]}^{\left[\frac{\Delta}{2}\right]} P(i)\right)}, \quad \textbf{(11)}$$

**with** $8 < \Delta < N_P$**.**

The proposed method has the following steps:

1. We apply Eq. (10) to calculate λ as a function of Δ up to a value $\Delta_{max} \leq N_P$. We plot λ versus Δ, and since we are interested in the convergence of λ [29], we focus on the rightmost part of the plot (for large Δ), where up to the last 10 points, i.e., $n = \Delta_{max} - 9, ..., \Delta_{max}$ [25], are sufficient to draw a conclusion. The value of $\Delta_{max}$ is determined in the tail of the laminar-length distribution according to the following criterion.
2. We quantify the previous step by calculating the distance of λ from the value $\lambda = 1$, which denotes an exact power law, by calculating:

$$Q_\lambda = \frac{1}{10} \sum_{n=\Delta_{max}-9}^{\Delta_{max}} (1 - \lambda_n)^2 . \quad (12)$$

The closer the $Q_\lambda$ is to 0, the more closely the distribution follows a power law. Thus, the position of $\Delta_{max}$ is determined where the quantity $Q_\lambda$ comes closest to zero.

3. Then, we produce the $R$ vs Δ plot using Eq. (11). From the convergence region of the plot, i.e., for $n = \Delta_{max} - 9, ..., \Delta_{max}$ a mean value, $\langle R \rangle$, is obtained for the quantity $R$.
4. We consider the test function $f(i) = c_f i^{-p}$, $i = 1, 2, .. \Delta_{max}$, where $c_f$ is a constant factor, and we solve Eq. (11) numerically for the given $\langle R \rangle$ value with respect to $p$, to estimate the value of the $p$ exponent corresponding to $\langle R \rangle$.

The power-law test function mentioned in Step 4 is used only when $Q_\lambda < order(10^{-3})$, which indicates an exact power law. Furthermore, if $p \in [1, 2)$, this indicates that the underlying system is in a critical state. If $order(10^{-3}) < Q_\lambda < order(10^{-2})$, this indicates that the laminar-length distribution under examination is close to a power law. In this case, in Step 4, we can use a truncated power-law test function $g(i) = c_g i^{-p} e^{-iq}$, $i = 1, 2, .. \Delta_{max}$, where $c_g$ is a constant factor, instead of a power-law test function $f(i)$for further investigation. We thus quantify a power law with an exponential correction. If the exponents are found to be $p \in [1, 2)$ and $q \approx 0$, this indicates that the underlying system is in a critical state. In cases where $Q_\lambda$ is greater than $order(10^{-2})$, there is no power-law distribution.

## Section 4: Application of the WL method to an EEG time series

The analysis following the presentation of the method was applied to an EEG time series from a healthy individual. The data were introduced by Andrzejak et al. [30] and have been analysed using statistical methods such as Tsallis entropy and Shannon block entropy in [31]. The set containing 100 single-channel EEG segments, each of 23.6 s duration, was used for this study. These are EEG recordings from a healthy individual. The segments satisfy the criterion of weak stationarity [30]. After 12-bit analogue-to-digital conversion, the data were written continuously to the disk of a data-acquisition computer system at a sampling rate of 173.61 Hz. The band-pass filter settings were 0.53–40 Hz (12 dB/oct.).

We previously conducted a similar analysis using the initial version of the method [29]. The present work employs an improved version of this methodology. In our earlier study, we also analysed an EEG recording from an individual with epilepsy. In the present work, we focus exclusively on the healthy brain as a reference point, so that future work can compare it with neurological deviations from healthy behaviour.

Fig. 3 presents the results obtained using the algorithm described in Section 3.

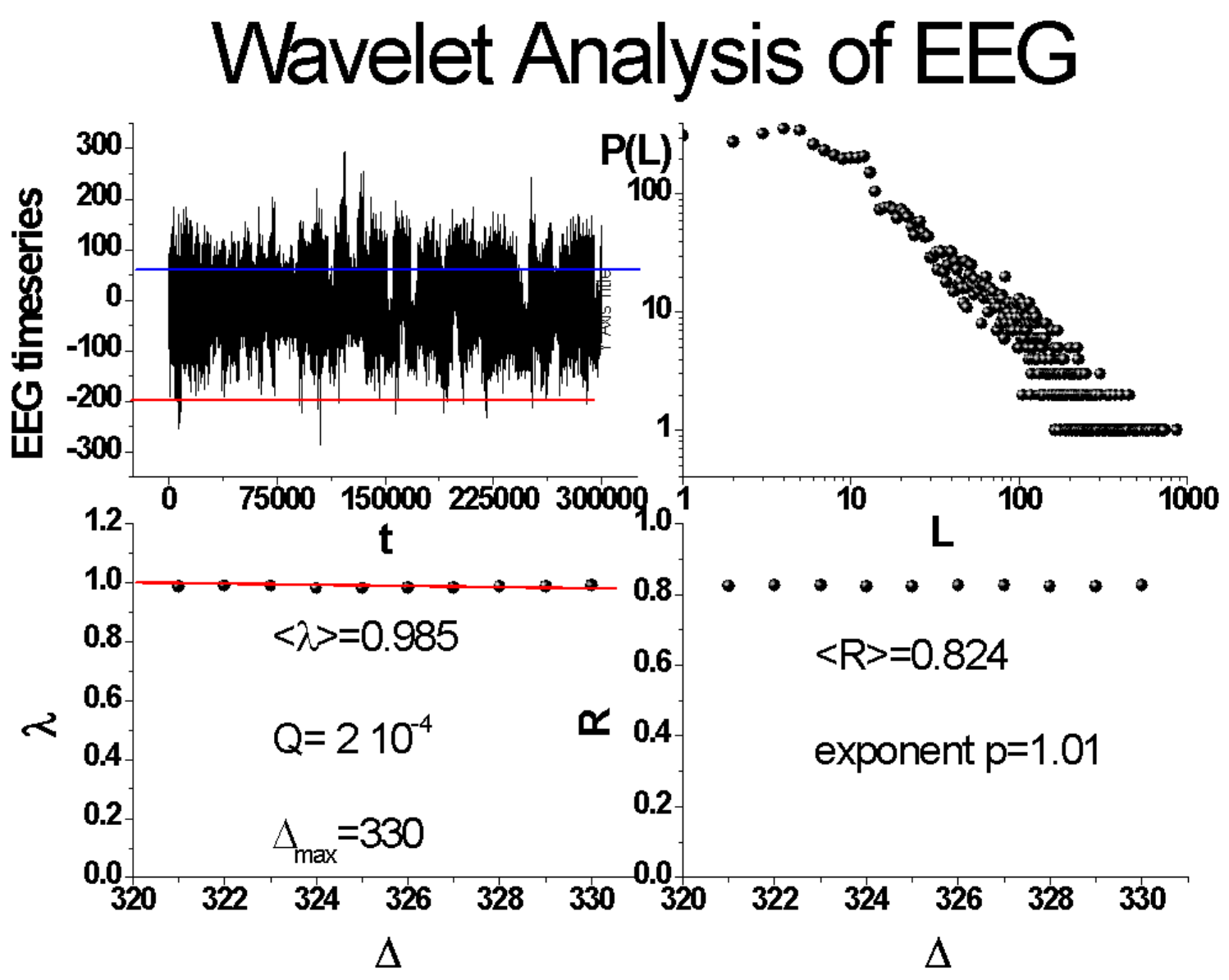

**Fig. 3. (a) The EEG time series, consisting of 300,000 points. The red line represents the fixed point (here, -200), and the blue line represents the endpoint, determined to be 60. Thus, the laminar region is [-200, 60]. (b) The distribution of laminar lengths within this region. Linearity in the log–log distribution is examined in its tail, i.e., for L >> 1. (c) The algorithm presented above determines λ from Eq. (10) and the quality factor Q from Eq. (12). The last 10 values of λ lie almost exactly on the line λ=1, with a mean of <λ>=0.985. The scale at which the optimal quality factor Q is found is Δ=330. (d) The values of R are obtained from Eq. (11). The numerical solution obtained using the algorithm yields an exponent p=1.01 ∈[1, 2).**

This value indicates that the brain from which this EEG was recorded is marginally in a critical state; that is, it is functioning normally. Comparing these results with those obtained by applying the same algorithm to the 3D Ising spin model, we find p=1.03 [6] at T=4.49, where the resonance peak occurs within the hysteresis zone. This is very close to the value of 1.01 found for the brain using the same methodology. This is a striking result, as it demonstrates that resonance within the criticality hysteresis zone occurs in the brain, just as it does in the 3D Ising model. Resonance is a unique natural phenomenon in which a system absorbs the maximum possible energy, leading to the maximisation of all its parameters. In previous work, we demonstrated that resonance also emerges within the critical state of a p-n electronic junction, where the control parameter is frequency rather than temperature. We therefore consider a correspondence between temperature and frequency as control parameters.

# Section 5: Domain walls in the brain

Brain neurons are electrically charged, owing to protein ion pumps (typically for $K^+$ and $Na^+$) that transport ions across their membranes. Consequently, neurons constantly exchange ions with the extracellular environment. Ions with like charges repel one another; thus, when many ions are expelled from numerous neurons simultaneously, they exert forces on neighbouring ions, which, in turn, push their neighbours, and so on, generating a wave. This phenomenon is known as "volume conduction". When this wave of ions reaches the EEG recording electrodes, electrical conductivity allows it to be transmitted to them. An EEG therefore records the summed, synchronised activity of thousands of neurons that share a similar spatial orientation. If the neurons in a region lack a similar spatial orientation, their ions fail to align in a way that produces detectable waves. It follows that, whereas the criterion for "domain walls" in magnetic materials is the orientation of the local spatial magnetisation, as determined by the spins, the corresponding criterion in the brain is the local spatial orientation of the neurons in a given region.

It is also well established that EEG activity exhibits oscillations at various frequencies. Many of these oscillations have characteristic frequency ranges and spatial distributions and are associated with different states of brain function. They represent synchronised activity within a neural network. In other words, when the brain is in a critical state, resonance emerges at a specific frequency of the electromagnetic (EM) field generated by moving ions. This implies, for instance, that various regions of the brain function as transmitting antennas, optimising their operations through resonance. We conclude this section by observing that an ionic wave originating from a neuron and sweeping across a local neural network (a "domain wall") is precisely analogous to a "spin wave" [11] in a ferromagnetic material in a critical state. In both cases, the wave propagates through the medium, carrying spatio-temporal correlations from one end to the other. These correlations compel the constituent elements—spins in magnetic materials or neurons in neural networks—to align in a common orientation within the local region.

## Section 6: Autocorrelations in the 3D Ising model and the brain

The autocorrelation function of a time series is useful because, among other things, it provides a qualitative measure of the temporal memory of the system generating it. In general, a long-memory correlation function is a strong indication that the time series is generated by a critical dynamical system, owing to its invariance under changes in spatio-temporal scale. We use it to investigate criticality in the brain.

The autocorrelation function *C(m)* for a time series $\left\{x_1,\ x_2,\ x_3, \ldots\ldots x_n\right\}$ is defined as:

$$C(m) = \frac{\sum_{i=1}^{n-m} (x_i - <x>)(x_{i+m} - <x>)}{\sum_{i=1}^{n} (x_i - <x>)^2} \qquad (13)$$

where $m<n$ and

$$< x >= \frac{1}{n} \sum_{i=1}^{n} x_i$$

Qualitatively, autocorrelation curves in critical systems take the form of a nearly straight line, typically with fluctuations and a slight downward slope, extending along the "memory time" axis. Quantitatively, in a normalised plot with a maximum autocorrelation value of unity, the closer the line remains to unity, the greater the degree of memory.

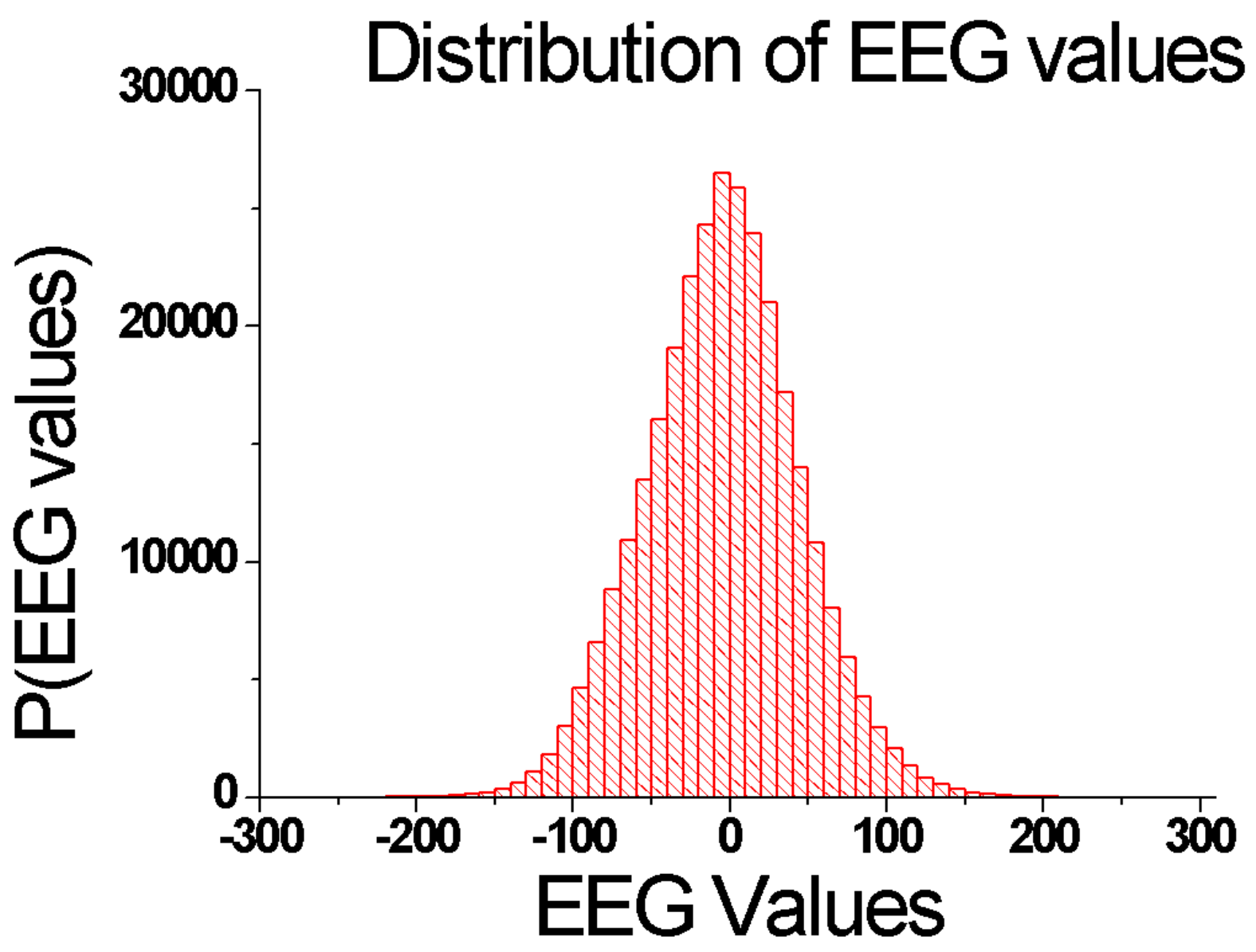

**Fig. 4. The distribution of EEG values. The distribution is symmetric about 0, meaning that the numbers of positive and negative values are equal. Consequently, the positive component of the autocorrelation is cancelled by the negative component (anticorrelation), resulting in a curve that fluctuates around zero.**

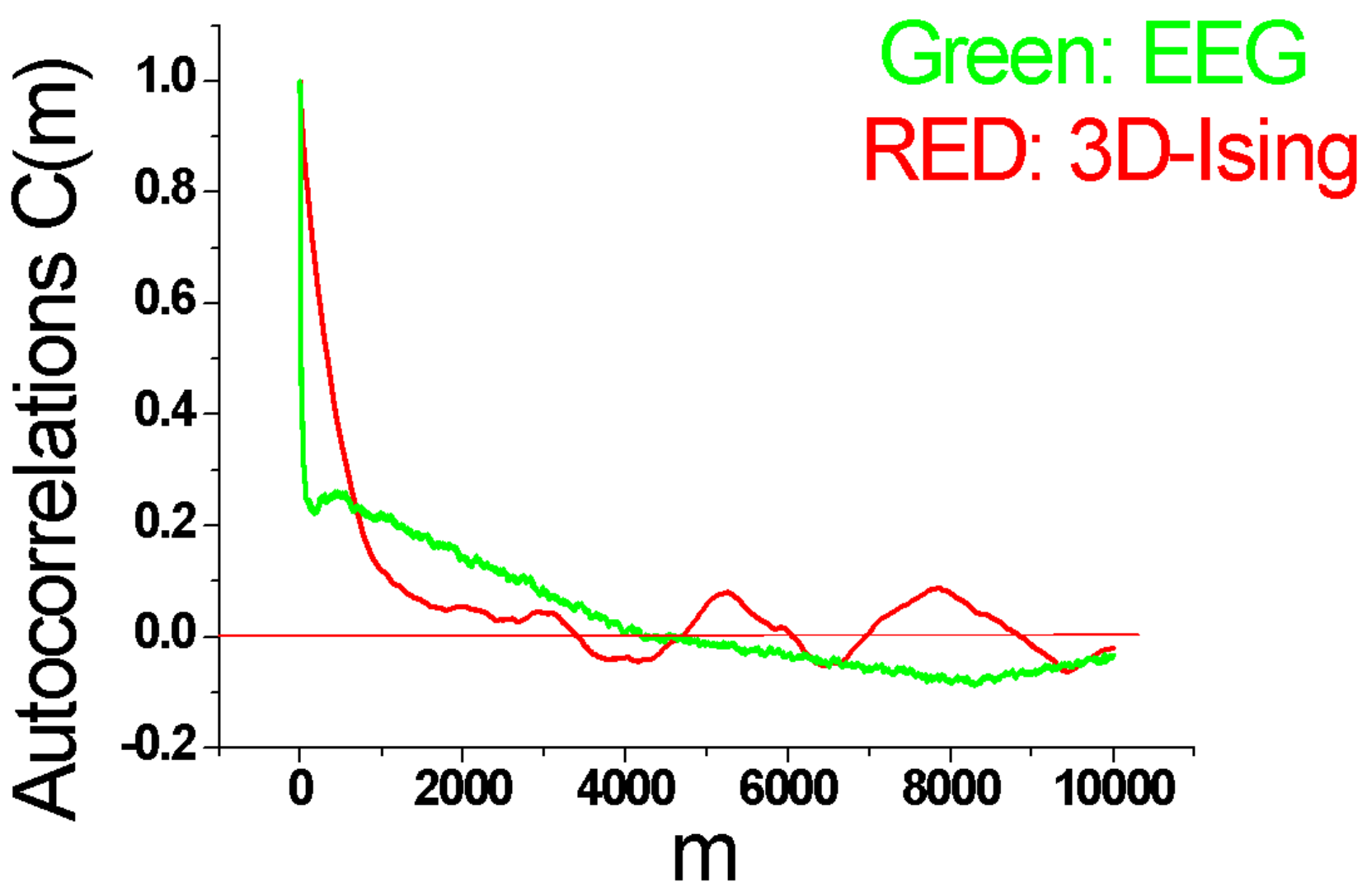

**Fig. 5. The green curve shows the autocorrelation function of the EEG time series up to m=10,000. The result shows the presence of anticorrelations. As noted above, information about the critical state has been cancelled by anticorrelations, as a consequence of symmetry about zero. For comparison, the red curve shows the autocorrelation of the 3D Ising model time series at the resonance temperature T=4.49. The results are similar to those for the EEG, with anticorrelations present. The reason is the same as in the EEG case: symmetry about zero, as illustrated in Fig. (7) in Appendix A for the 3D Ising model. It is immaterial that the Ising model exhibits two lobes while the EEG exhibits only one; the determining factor is that both are symmetric about zero.**

Therefore, to capture the autocorrelation functions, we must eliminate the anticorrelations by restricting the analysis to the positive component of the symmetry. This is possible only because of the symmetry about zero. Another way to eliminate anticorrelations is to take the absolute values of the time series when the symmetry is perfect.

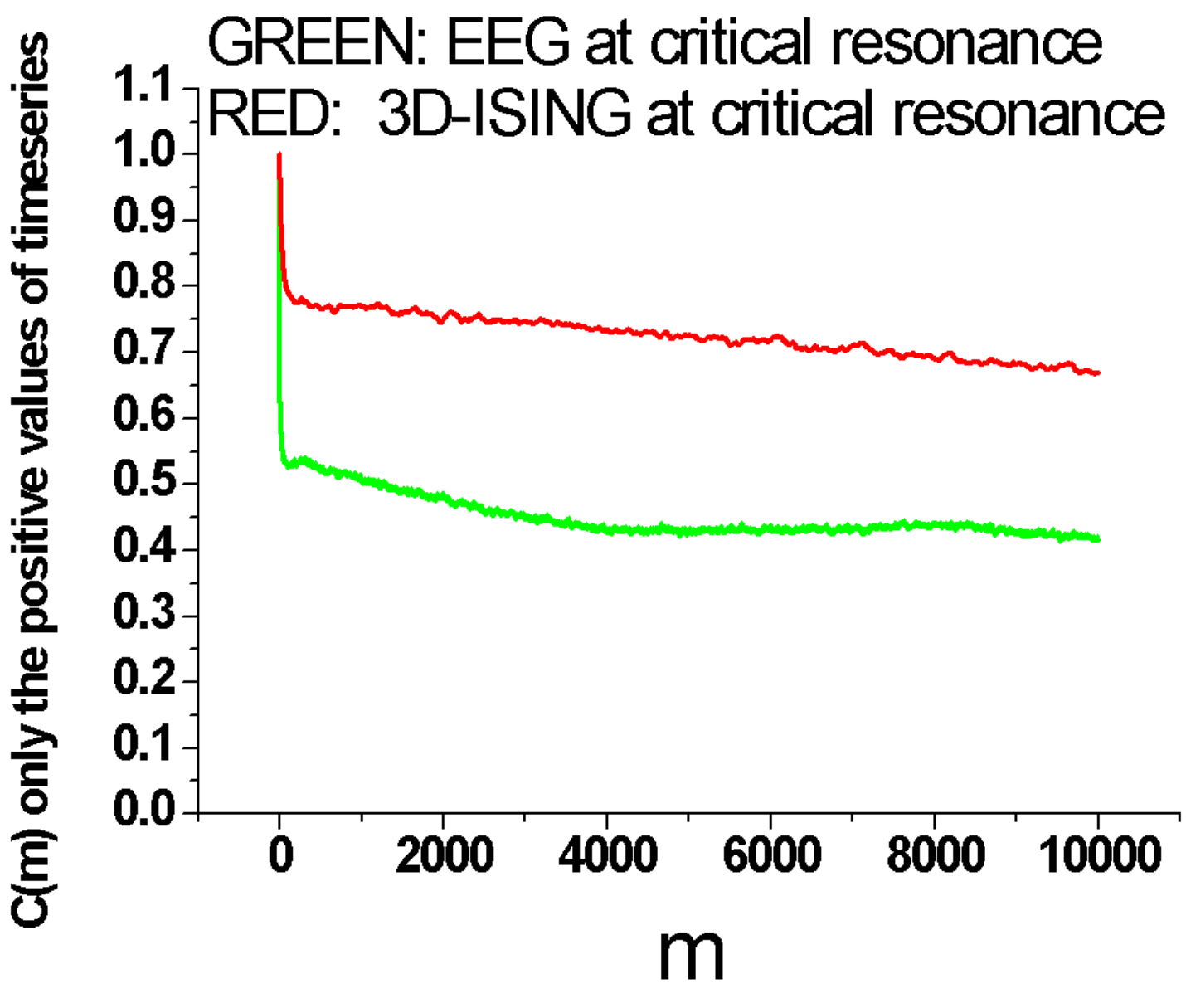

**Fig. 6. The red curve shows the autocorrelations for the portion of the EEG time series corresponding to positive values. A very high degree of temporal memory stability is observed (m=10,000). The green curve shows the autocorrelation of the 3D Ising spin model, likewise for positive magnetisation values and in the state of critical resonance, which also exhibits memory stability. In general, the EEG autocorrelation, which captures resonance in the critical state, follows the same pattern as that of the 3D Ising model at critical resonance.**

In this section, we presented similarities between the dynamic behaviour of the 3D Ising spin model—a prototype for the properties of ferromagnetic materials in a critical state—and that of the healthy brain, viewed as a critical system.

## Section 7: The Prime Number Algorithm (PNA) method

A major advantage of the Ising model is that it permits a binary approach, since its spins have only two possible orientations: +1 and -1. In a real magnetic material, however, all orientations are present. The Ising model effectively reduces these orientations to two values. This provides a coarse-graining method for generating symbolic dynamics based on two symbols: positive magnetisation values correspond to +1, and negative values to -1.

In our work [6], we applied coarse-graining at the pseudocritical temperature of the 3D Ising model. The distribution of magnetisation values is shown below.

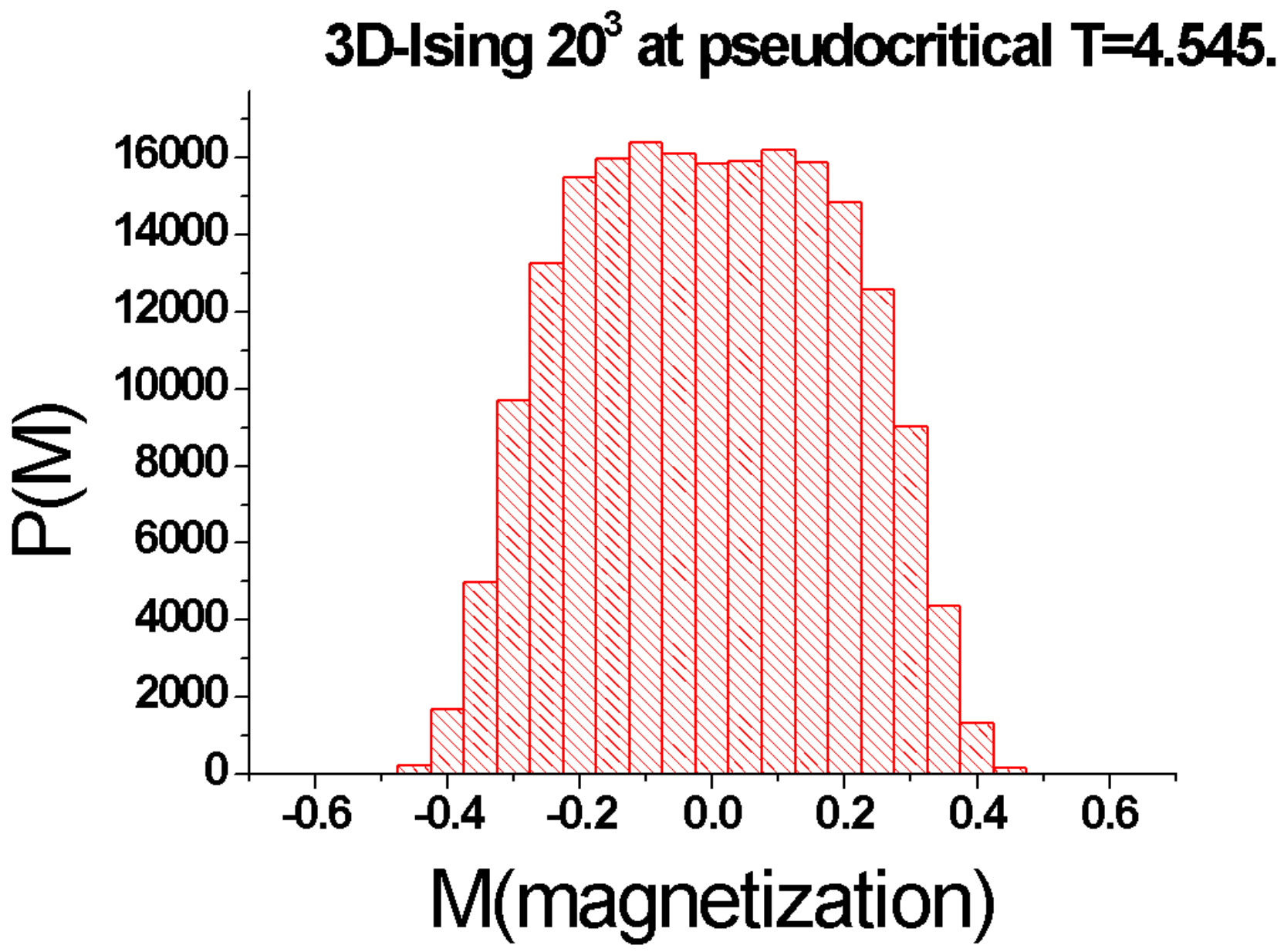

**Fig. 7. The distribution of magnetisation values for the 3D Ising spin model at the pseudocritical temperature T-4.545 (the beginning of the hysteresis zone). The distribution is symmetric about zero.**

First, we transform the magnetisation time series that produced the distribution above into a symbolic time series, assigning -1 to all negative magnetisation values and +1 to all positive values. In this way, we obtain the corresponding two-symbol dynamics.

In our work [1], we conducted a numerical experiment using a series of identical circular rings with a common axis passing through their centres. When an electric current of constant magnitude but varying direction (clockwise or anticlockwise) flows through this system, the Biot–Savart law gives a magnetic field within the cylindrical structure whose value varies from point to point as a classical field. The figure below illustrates this device according to the PNA methodology.

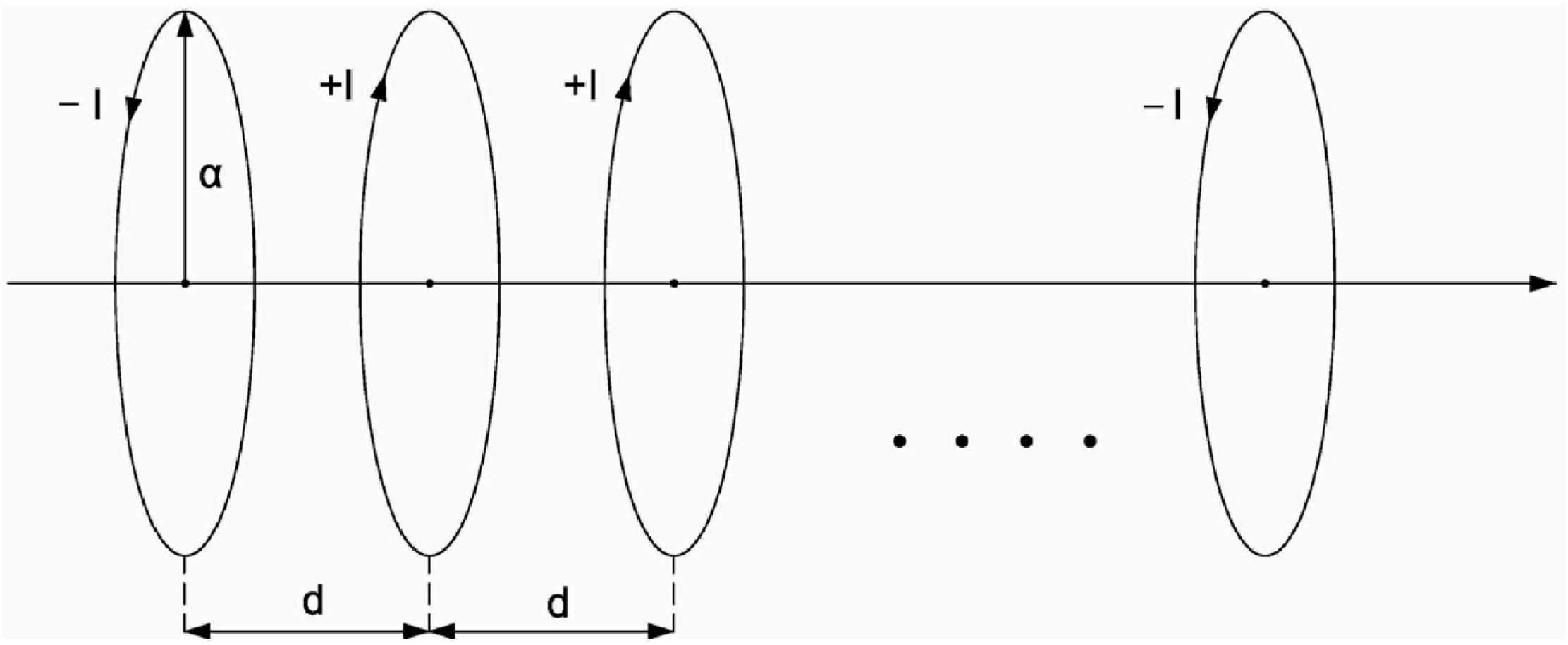

**Fig. 8. The PNA device. The ring radius is *a*, and the distance between consecutive rings is *d*.**

However, as we decrease the radius α while increasing the length d, specific zones of magnetic field values emerge, alternating with empty regions. Beyond a certain point—specifically, when the ratio of the ring radius α to the distance d, i.e., α/d, is of the order of 0.1—the magnetic field becomes quantised into two central values of $\pm$ 0.5 [1]. Four additional quantum levels appear on either side of these values, giving 6 levels in total. Each central quantum level accounts for 25% of the quantum values, so the two central levels together account for 50% of the total. The remaining 50% is distributed across the other 4 quantum levels. As we demonstrate through detailed calculations in [1], this transformation of the classical magnetic field into a quantised field is achieved using the set of prime numbers from theoretical mathematics.

As shown in Fig. (7), the distribution of magnetisation values forms a single lobe symmetric about zero and includes both positive and negative values. The same distributional form is observed for the EEG data (Fig. 4). We therefore apply PNA to the EEG time series. Its criticality has already been addressed in the section on the wavelet method. We must now ask whether applying PNA to the EEG time series yields magnetic field quantisation into 6 levels, as in the 3D Ising model. For this question to be meaningful for a real physical system such as the brain, the quantisation must be more than the result of a mathematical algorithm: the algorithm itself must be implemented by mechanisms in nature. In the case of the brain, this is arguably the fundamental question distinguishing artificial intelligence from genuine, natural intelligence.

In previous work [1], we applied PNA to various systems whose time-series values exhibit a two-lobe distribution. This applies whether the lobes are in direct contact, as at the pseudocritical temperature of the 3D Ising model (Fig. 7) or in the distribution of EEG values (Fig. 4), or are separated by a gap, as in the distributions shown in the figure in Appendix B.

# Section 8: Quantum computers and application of the PNA method to the brain

The brain does more than sustain resonances and control the body's organs: it also generates what we call our mental world—our thoughts, ideas, emotions and, collectively, consciousness. This is the most challenging aspect of the research: how does the tangible matter of the brain give rise to our "immaterial" emotional world, with all that it entails? This remains the greatest scientific question in the field, and progress towards answering it has been negligible. Under the right conditions, all the material systems around us can reach a critical state, yet none produces emotions. Relying on scientific explanations does not imply that science cannot answer such questions; rather, it must extend its reach into the hidden realm of our "immaterial" consciousness. This has happened repeatedly in the history of science over at least the past 300 years. For example, Newton's law of gravitation is not incorrect; Einstein extended it through the General Theory of Relativity to address problems that Newton's law could not solve. PNA is an algorithmic approach. Its objective is to calculate the dynamics of a symbolic time series by analysing the distribution of "islands" of quantised values across two central quantum levels. For a ratio $\alpha/d=0.1$, these correspond to quantised values of $\pm$ 0.5. It is applicable to any time series whose value distribution exhibits two distinct lobes. A two-symbol time series is constructed externally by the operator through coarse-graining: one symbol is assigned to the values in one lobe and another to those in the other. When positive and negative values are intermixed, as in the time series from the 3D Ising model, +1 is assigned to positive values and -1 to negative values. There is a significant distinction between the present application and other algorithmic applications of PNA: here, the algorithmic steps are performed by the brain itself rather than by an operator.

A quantum computer operates using the superposition of two states, represented as qubits (0, 1), where 0 denotes an "open" switch and 1 a "closed" switch. The switch therefore exists in two states simultaneously, being both open AND closed, contrary to our conventional logic. This offers a measurable advantage: using qubits (0, 1), the system performs parallel computations, making it significantly faster, with the effect amplified as more qubits are added. This phenomenon is not observed in the human brain, whose computational capacity is infinitely smaller than that of a conventional computer, let alone a quantum computer. In PNA, when the ratio $\alpha/d$ within the ring system approaches zero, only the two central quantum states remain: ±0.5. PNA therefore allows us to propose that the human brain could potentially function as a quantum computer using qubits defined by the states (-0.5, +0.5). In reality, however, it cannot be a quantum computer for the following reasons: (a) The brain is a finite system, and its smallest length scale is of the order of Å, where quantum phenomena just begin to emerge, making the ratio $\alpha/d \to 0$ unrealistic. (b) A quantum computer collapses through a phenomenon known as decoherence when it interacts with its environment. Since the brain is an open system, interaction with the environment is inherent.

(c) A quantum computer encompasses all possible quantum states with varying probabilities, effectively closing the gaps between the "islands" of quantum values at any given level. A distribution of island lengths would therefore be virtually impossible, because no gaps would separate the islands. Consequently, the scaling laws and, by extension, the critical state break down, specifically disrupting the resonance within the hysteresis zone. Without the resonance of neurons towards a common orientation, the neural networks would fall into total chaos and would be incapable of supporting structured thought. PNA functions as a "semi-quantum computer" because it uses two central quantum states, for example ±0.5, while employing a logical "superposition" in the form of an OR between the states -0.5 and +0.5 (see the PNA algorithm in [1]). It operates according to the following instruction: measure the length of the islands (clusters) of magnetic field values, whether they lie at the first central quantum level (+0.5) or the second (-0.5), a process consistent with the reality we perceive. In other words, it accounts for both states simultaneously, but uses a disjunctive "superposition" (an OR operation) rather than an exclusive superposition (an AND operation). This approach can be extended to the other four quantum levels. Our advantage over a standard quantum computer therefore lies in the freedom to select combinations of states, a capability essential to generating complex emotional thought.

Returning to biological matter, our brains contain microscopic structures on the scale of Å that share the geometric characteristics of the ring system in the PNA model. Microtubules are hollow, cylindrical protein polymers that form the structural framework and transport network within neurons. From biology, we know that they have a diameter of 250 Å and a radius of 125 Å. There are at most 17 such molecular rings in an entire microtubule, spaced 80 Å apart. There are therefore 17–1=16 intervals, giving a microtubule length l of 80×16=1280 Å. Since the number of "rings" in a microtubule is much smaller than that in a sequence within the PNA device, we consider the entire microtubule length, l=1280 Å, for comparison. The radius-to-length ratio is therefore 125/1280=0.097 ≅0.1. The biological microtubule therefore satisfies both the structural and dimensional requirements of our experimental ring-based set-up (α/d=0.1), which produces 6 quantised magnetic fields. Essentially, we have assumed that a microtubule has two rings, at its entry and exit, such that its total length l corresponds to the distance d=1280 Å between them. Since thousands, or even hundreds of thousands, of microtubules are arranged along the axon within each neuron, depending on its length, each neuron ultimately contains thousands of such rings, as in a PNA device, spaced d=1280 Å apart.

As noted in Section 7, a PNA device requires only an α/d ratio of the order of 0.1, rather than exactly 0.1. This therefore accommodates microtubules of various lengths, not only those measuring 1280 Å, provided their ratios remain within the same order of magnitude, i.e., $10^{-1}$. In these cases, the results remain unchanged: we still observe the same six quantum levels and the same distribution of "islands". Although the specific positions of the levels may shift, their underlying symmetry remains constant.

We can therefore apply PNA using exactly the same procedure as for the 3D Ising model at the pseudocritical temperature.

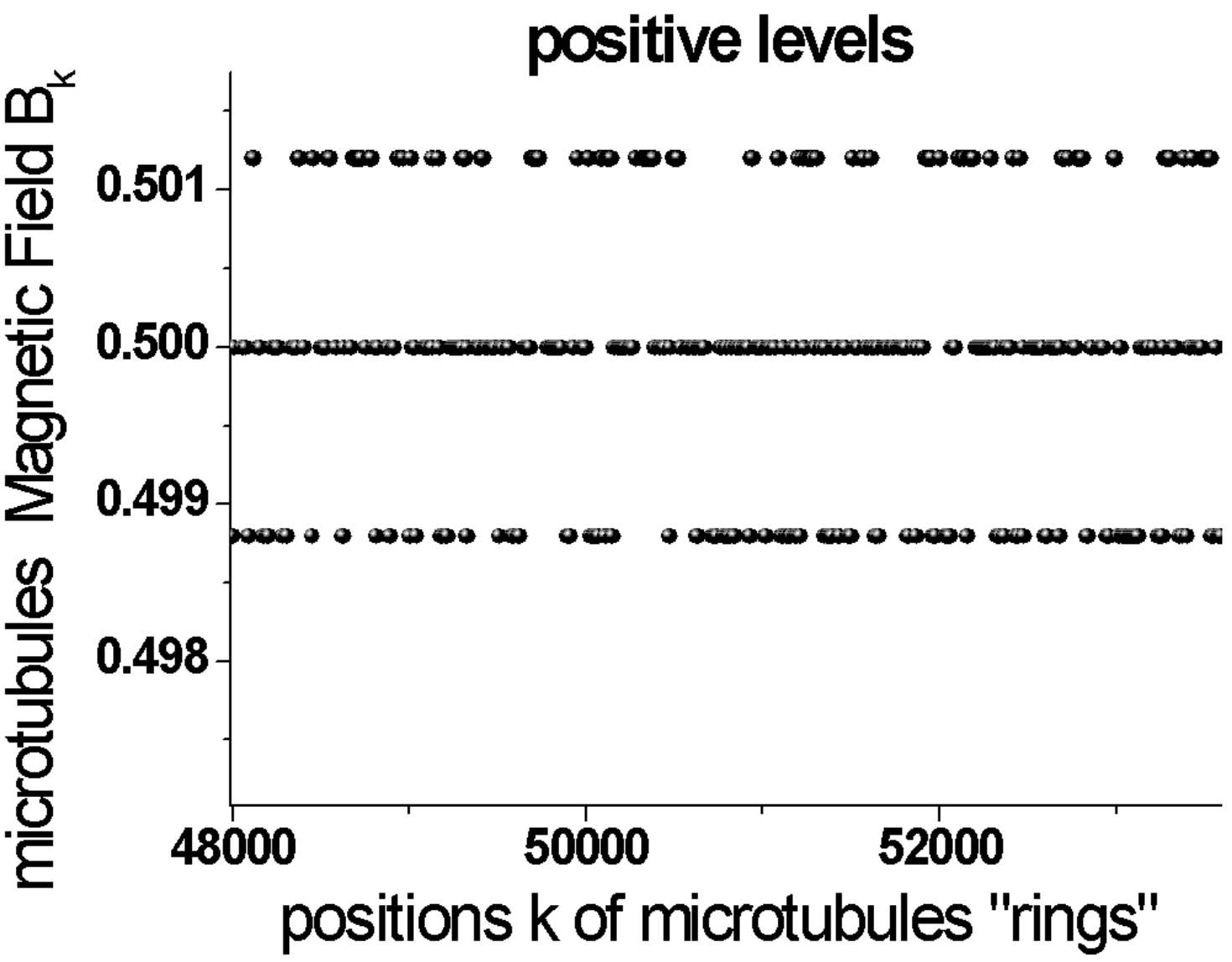

**Fig. 9. The three positive quantum levels of the magnetic field of the microtubules. Corresponding negative levels occur at symmetric positions. Clusters of magnetic field values are clearly visible, and the horizontal axis shows their positions. The gaps between clusters at each level are essential for determining the cluster distribution. The figure represents a randomly selected short segment of the EEG time series.**

**Remark: The diagram gives the impression that some positions correspond to more than one field value. This is not the case: further thinning of the points reveals that each position corresponds to exactly one magnetic field value.**

# Section 9: A proposed mechanism of thought based on the PNA method

How is the PNA algorithm implemented by the brain? The brain's electrical charge is maintained by billions of neurons, which are electrically charged because they pump ions across their membranes. Neurons constantly exchange ions with the extracellular environment. This process generates action potentials (spikes) that propagate along the axon. As noted in Section 5, ions with like charges repel one another. When many ions are expelled from numerous neurons simultaneously, they exert forces on neighbouring ions, which, in turn, push their neighbours, and so on, creating a wave. As EEG recordings show, this wave exhibits pronounced fluctuations between positive and negative values. When these electrical oscillations within neurons reach the microtubule region, the following occurs: depending on their positions within the wave, the negative and positive electrical currents pass across the protein surface of the microtubule cylinder. This implies a sequence of clockwise and anticlockwise flow directions for positive and negative ions, similar to the ring system in the PNA device. Furthermore, the distribution of wave values shown by the EEG is symmetric about zero. The positive values are therefore approximately equal in absolute magnitude to the negative values; under perfect symmetry, this equality holds exactly. The positive and negative currents traversing the microtubules consequently have approximately equal magnitudes. This provides another point of consistency with the PNA device. Since the flow of an electrical wave generates a surrounding magnetic field, quantised magnetic field levels can be expected to emerge in the microtubule region, as in the PNA device.

Considering the equivalence of the transformation between the magnetic field before entering the microtubule region and the field upon exiting, and given that the critical dynamics remain invariant under this transformation [1,25], we conclude that the information carried by the action potential from external and internal stimuli is transferred into these quantum levels. Through the OR process, networks of quantum levels across millions or even billions of neurons possess immense combinatorial freedom, providing a rich substrate for thought. This combinatorial freedom of information, inherent in neuronal microtubules, shapes neural network patterns within the brain, unconstrained by the rigid logic of an "AND" operation. These patterns evolve continuously over time. Some may become "frozen" in time, forming long-term memory. The OR mechanism thus explains why individuals exposed to the same stimuli can develop different thoughts and, consequently, distinct forms of consciousness, as their microtubules select different pathways. A critical state is essential for thought and consciousness: only within this state can "domain walls" communicate, scaling laws encompass all correlation lengths, and resonances establish a common orientation among local neurons and their microtubules. Far from criticality, connections between microtubules would instead be disorganised, resulting in a lack of coherent thought. The image below shows neuronal microtubules.

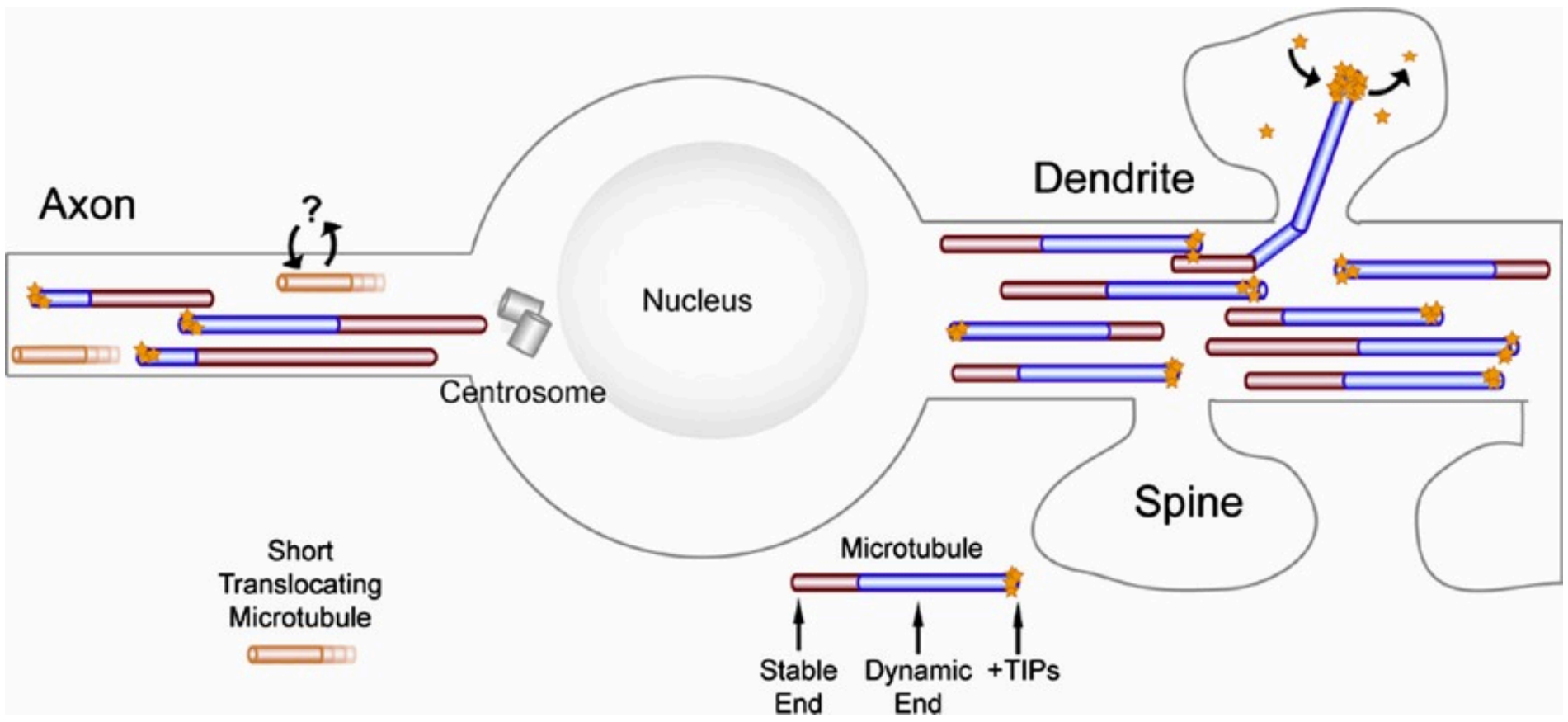

Image of microtubules from reference [32], by Eric W. Dent and Peter W. Baas, published in the Journal of Neurochemistry.

# Section 10: Communication between quantum levels

To summarise, information about an external or internal stimulus is transmitted to the quantum levels of the magnetic field within neuronal microtubules. This information is encoded as a distribution of "islands": all points within a given level share the same quantum value (e.g., ±0.5 for the two central levels), with gaps between the islands. For the reasons explained above, these quantum values do not exhibit quantum superposition; instead, they exhibit logical superposition through the OR operator.

It remains to explain how the OR operation is implemented between quantum levels. Computationally, it is implemented through appropriate commands within an algorithm. How, however, does this communication occur naturally within the brain? We address this question by returning to Fig. 1. Following spontaneous symmetry breaking (SSB), the resulting energy wells do not coexist; rather, either the positive or the negative well is present. However, precisely at SSB, the distribution of magnetisation values (Appendix A) exhibits two symmetric lobes poised to separate. At this stage, we are still within the hysteresis zone. As is known from the physics of critical phenomena, the energy wells in Fig. 1 communicate through kink–antikink solitons [19,20]. Solitons have a stable form that remains intact even when they interact. They propagate through space and time as a wave of energy (a solitary wave). Since they have mass, we regard them as material particles (quasiparticles). The figure below illustrates one such kink soliton and its energy density [19].

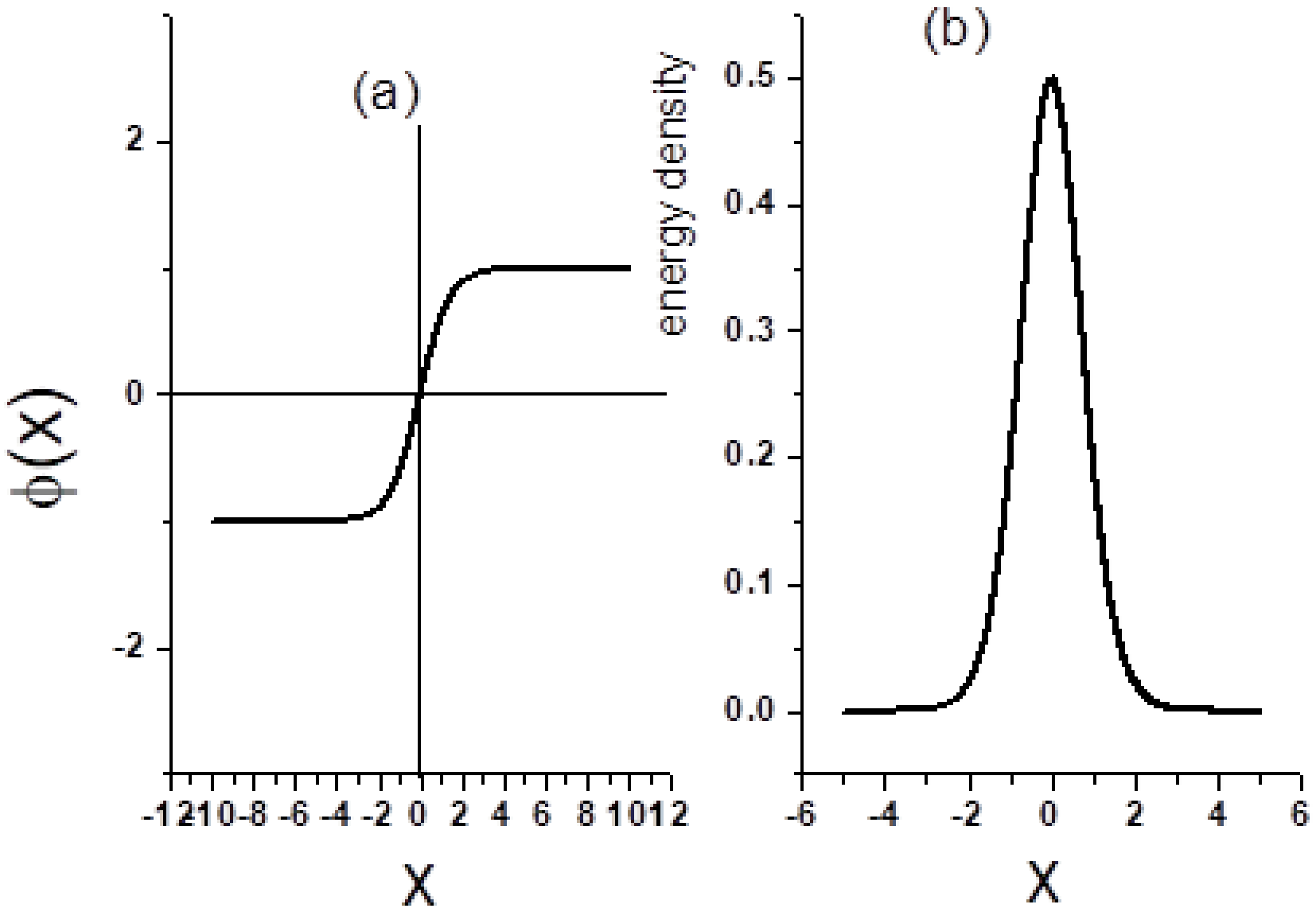

**Fig. 10. (a) The soliton propagating in spacetime x (a solitary wave) is a kink-type soliton. (b) The soliton is a quasiparticle and therefore possesses energy and mass. Its energy density is shown in the figure. The field φ(x) is the field of the $\varphi^4$ theory (Eq. (7)).**

Furthermore, when the conditions for PNA, as observed in microtubules, are met, the classical magnetic field is transformed into a six-valued quantum magnetic field. All these processes occur within the hysteresis zone and the SSB region. The classical energy wells in Fig. (1) are therefore replaced by 3 positive and 3 negative quantum levels. Communication between these levels occurs through the OR operation, allowing solitons to move back and forth among them. As noted above, this OR mechanism enables nature to ensure a critical state and, more importantly, resonances of microtubules across all neurons within a neural network, specifically in the synaptic regions. This also implies that the solitons mediating communication have enormous combinatorial freedom in their quantum values, not only within a single neuron but across the hundreds of thousands of neurons in the network.

In nature, communication between quantum levels takes the form of excitations and de-excitations. One example is the nuclear shell model, which explains the so-called "magic numbers" [33,34]. A confirmation of the magic-number shell model was recently achieved [35] using prime numbers, as in PNA.

As we demonstrate in Appendix B, the hysteresis zone contains exotic matter consisting of tachyons and solitons, both of which have imaginary mass. Owing to this imaginary mass, these particles travel faster than light. If the complete process of thought generation begins when a stimulus triggers spikes within a neuron, which then propagate along the axon towards the synapses, the combinatorial selection of quantum levels constitutes its final stage. The stages preceding the quantum levels within the microtubules proceed at speeds infinitely lower than the speed of light. However, the final and most critical phase in generating thought—the coupling of the quantum levels of the magnetic field—is mediated by superluminal, imaginary-mass solitons. In other words, the selection of couplings between quantum levels is nearly instantaneous.

## Section 11: Discussion and conclusions

The brain exhibits an extraordinary physical property, hitherto unknown: the microtubules within its neural networks can quantise the magnetic field generated inside them. As discussed above, this field is associated with the flux of positive and negative ions passing through neurons and carrying information and stimuli. In physical terms, the healthy brain is in a state of constant resonance because it operates in a critical state**. This resonance and magnetic field quantisation are the keys to cognitive functions,** for the following reasons. (a) Resonance imposes order on the orientation of neurons and their microtubules, much as spins align in a ferromagnetic material at its critical temperature, creating robust correlations across all scales. (b) Quantisation restricts the magnetic field to six discrete quantum states drawn from the infinite range of values present in a chaotic magnetic field. This produces structured correlations among neurons within networks, forming organised structures that constitute thought. These correlations represent a form of semi-quantum entanglement arising from the brain's resonance state. Here, the quantum principle of superposition of quantum states, based on an AND operator that transcends human logic, does not apply.

Instead, a combinatorial superposition based on logical OR is present. The OR operator permits numerous free combinatorial choices, facilitating complex thought. It also allows gaps within quantum levels, producing a distribution of "islands" of values—a condition that is both necessary and sufficient for a critical state and, by extension, resonances. **Thus, although the brain is not a quantum computer, it encodes information at quantum levels, whether received externally or retrieved from memory.** Communication between microtubules, mediated by complex-mass solitons, enables thought networks to be organised at speeds exceeding that of light. This does not imply that thought itself "moves" faster than light, because most of the sequence, from the initial stimulus to quantisation, occurs at very low speeds. These networks, formed through interactions between microtubules via solitons acting as material particles, constitute the physical realisation of mental functions such as thought. It should not surprise us that matter can move faster than light: such a phenomenon occurred in the earliest moments of the Universe, during the inflationary period, when prime matter underwent massive expansion at velocities far greater than the speed of light.

We consider that this work opens a window onto the perennial question of how a material object such as the brain produces the "immaterial" world of consciousness.

## Appendix A: The 3D Ising spin model

To study phase transitions computationally, statistical physics employs Z(N) spin systems, the most prominent example of which is the 3D Ising model described below.

For a Z(N) spin system, spin variables are defined as: $s\left(a_j\right) = e^{i2\pi a_j/N}$ (lattice vertices $j = 1, 2, \ldots j_{max}$, with $a_j = 0, 1, 2, \ldots N - 1$. For N=2, the spin has two possible states, +1 and -1. For N=3, there are three possible spin states$\left| -\frac{2\pi}{3} \right\rangle$, $|0\rangle$, and $\left| +\frac{2\pi}{3} \right\rangle$ $and\ so\ on$ [36]. Such a thermal system can be studied using the Landau approach to critical phenomena [11]. In general, the Ginzburg–Landau (G-L) free energy can be described as a polynomial function $U(\phi)$of the order parameter $\phi$ [11]:

$$U(\phi) = \frac{1}{2}r_0\phi^2 + \frac{1}{4}u_0\phi^4 + \frac{1}{6}c_0\phi^6 + \frac{1}{8}s_0\phi^8 + \dots \qquad (1A)$$

In the symmetric phase, all coefficients in Eq. (1A) are positive. This equation can describe first- and second-order phase transitions by retaining terms up to $\phi^6$ or up to $\phi^4$, respectively.

The 3D Ising model is a lattice model of ferromagnetism in statistical mechanics [11,19,20]. It successfully describes the second-order equilibrium phase transition, as well as more specialised topics such as spontaneous symmetry breaking (SSB) of the $\varphi^4$ Ginzburg–Landau (G-L) free energy [11,20]. The Metropolis algorithm is an effective method for generating configurations in Ising models. At constant temperature, configurations are selected with Boltzmann statistical weights $e^{-\beta H}$, where $\beta = \frac{1}{KT}$ and H is the Hamiltonian of the spin system with nearest-neighbour interactions, which can be written as:

$$H = -\sum_{<i,j>} J_{ij} s_i s_j. \quad (2A)$$

For a $20^3$ lattice in three dimensions (the 3D Ising model), the critical temperature (or pseudocritical temperature for finite-size lattices) has been found to be $T_c = 4.545$ ($J_{ij}$ =1) [16]. A sweep of the entire lattice represents the algorithmic time unit. Spins take the values ±1. The quantity recorded in the numerical experiment using the Metropolis algorithm [37] is the mean magnetisation M, which plays the role of the order parameter. The trajectory generated is a time series of order-parameter fluctuations. A time series of length $N_{iter}$ can be written in discrete form with a time step equal to the algorithmic simulation time, i.e., one lattice sweep. The simulation in the present work was performed for $N_{iter}$=300000 sweeps.

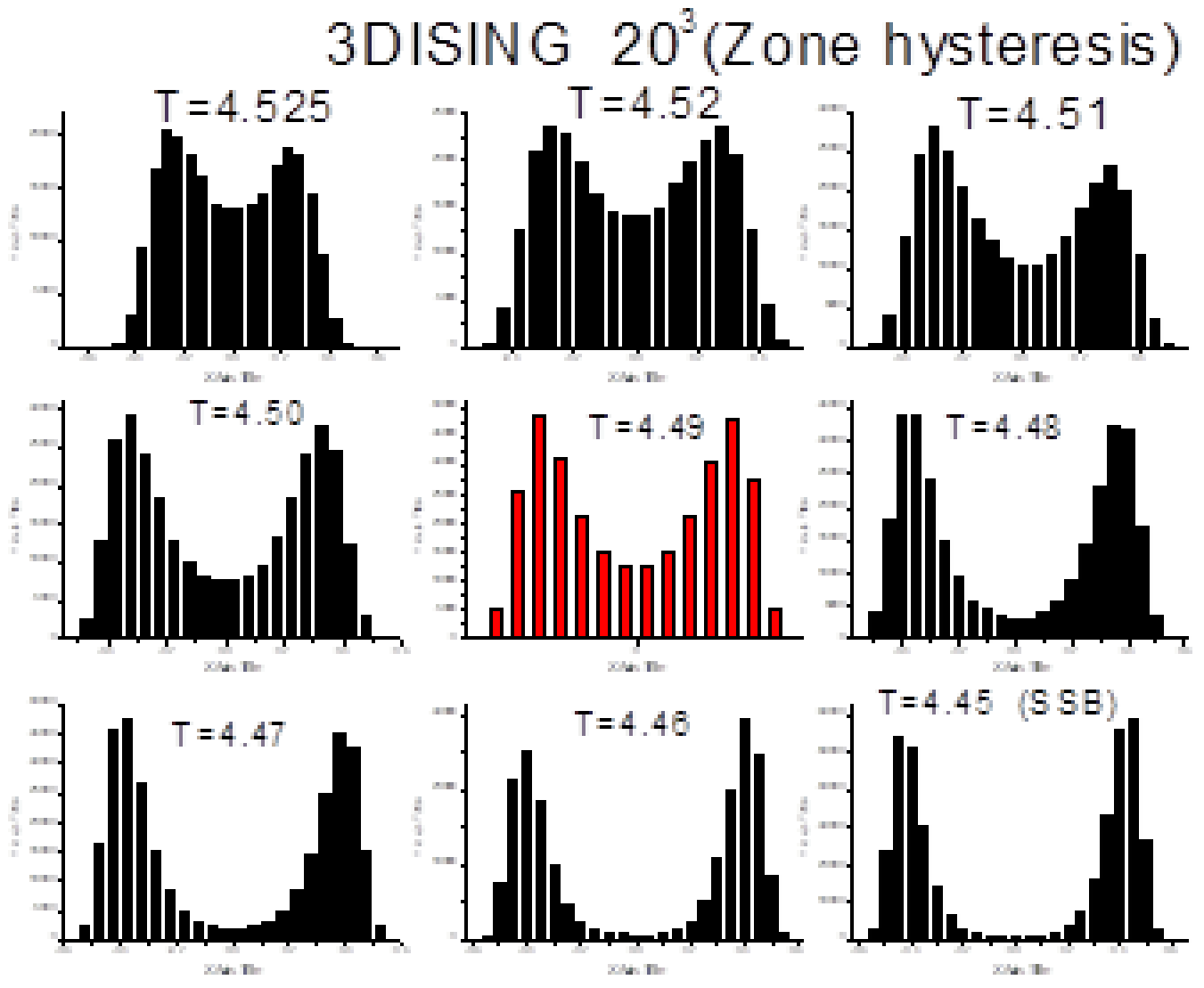

***Fig. A1. Nine distributions [6] within the hysteresis zone (see text), from T=4,525 to the spontaneous symmetry breaking (SSB) temperature $T_{SSB}$, for the numerical 3D Ising model on a $20^3$ lattice.***

## Appendix B: A discussion of exotic matter in the brain

So far, we have presented a series of proofs indicating that a healthy brain is in a state of perpetual criticality in physical terms. We have also observed consistency with the 3D Ising model in its critical state. We should therefore expect a hysteresis zone within the brain while it is in this state. This zone is dominated by "exotic matter". Below, we discuss this matter, which plays a fundamental role in mental functions or, in other words, in what we call the "soul".

We consider that investigations of the brain's "immaterial" aspect begin with a major error. Why do we label what we do not understand as "immaterial", and why do we cloak it in the metaphysical concept of the "soul"? As we shall see, matter is not limited to forms already known through direct or indirect experimental observation. It also includes

"exotic" matter, which has not yet been detected but is theoretically predicted by science.

As is known from the Klein–Gordon equation for scalar fields [19,20], the parameter $r_o$, which serves as the coefficient for the term $\frac{1}{2} r_o \phi^2$ , in the Ginzburg–Landau (G-L) equation for second-order phase transitions (Eq. (7)), corresponds to the square of the field's mass; that is, $r_o = m^2$ . This implies that, when the sign of $r_o$ changes from positive to negative, another significant change occurs: the mass m becomes imaginary, specifically, $m \rightarrow im$, where $i$ is the imaginary unit, defined as $i = \sqrt{-1}$. Particles with imaginary mass are called tachyons because, as follows from Lorentz transformations [20], they travel faster than light. Such particles have never been detected because the critical point is highly unstable; consequently, tachyons do not survive, at least in large systems. In finite systems, particularly microscopic ones, a hysteresis region emerges, as noted above, and extends to spontaneous symmetry breaking (SSB). Within this region, the critical point remains unstable, allowing the tachyon to survive. Furthermore, as we approach SSB and the distributions become distinct and cease to overlap, the two local minima of the potential (Fig. 1) emerge clearly. The unstable point disappears and is replaced, as a fixed point, by the two stable points corresponding to these minima. According to G-L theory, kink–antikink solitons (quasiparticles) then emerge and move between the two stable vacua. As established in standard theory, these solitons have a mass $M \sim m^3$, where the proportionality constant varies until it reaches its precise value at the SSB point. Since the mass parameter m is imaginary, i.e., m→im, the soliton mass is:

$$M \sim (i \cdot m)^3 \sim i \cdot i^2 m^3 \sim -im^3 \quad (17)$$

The soliton mass is therefore also complex. If we consider the coupling of two solitons, this coupling corresponds to a mass:

$$M_{coupling} = M \cdot M \sim \left(-im^3\right) \bullet \left(-im^3\right) \sim i^2 \bullet m^6 \sim -m^6 < 0. \quad (18)$$

In other words, negative mass emerges. This is another form of exotic matter that has not yet been discovered. Negative mass represents "anti-inertia", meaning that the gravitational force is repulsive: instead of attracting one another according to Newton's laws, two such bodies repel each other. The fact that negative mass, like complex solitons or tachyons, has not yet been detected does not mean that these forms of matter do not exist; rather, it indicates that our technology has not yet reached a sufficiently advanced level.